\documentclass[conference,page-numbering]{IEEEtran}

\usepackage[english]{babel}
\usepackage{semantic}
\usepackage{amssymb}
\usepackage{syntax}
\usepackage{listings}
\usepackage{textcomp}

\begin{document}
\title{Transformation of XML Documents with Prolog}

\author{\IEEEauthorblockN{Ren\'{e} Haberland \qquad Igor L. Bratchikov}
  \IEEEauthorblockA{Saint Petersburg State University\\
  Saint Petersburg, Russia}
}

\maketitle
\thispagestyle{plain}
\pagestyle{plain}
\pagenumbering{gobble}

\begin{abstract}
Transforming XML documents with conventional XML languages, like XSL-T, is disadvantageous because there is too lax abstraction on the target language and it is rather difficult to recognize rule-oriented transformations. Prolog as a programming language of declarative paradigm is especially good for implementation of analysis of formal languages. Prolog seems also to be good for term manipulation, complex schema-transformation and text retrieval.

In this report an appropriate model for XML documents is proposed, the basic transformation language for Prolog LTL is defined and the expressiveness power compared with XSL-T is demonstrated, the implementations used throughout are multi paradigmatic.
\end{abstract}

\section{Introduction}
Transforming XML documents means to build a target XML document from a source XML document
by rebuilding former parts and including new parts by querying external data.
Declarative approaches on transforming XML documents already exist in the functional paradigm. XSL-T
and XQuery represent those XML transforming languages that hold semantic equivalence by avoiding side effects. Alternatively imperative approaches like \cite{Xact} have been proposed. Both approaches lack either from abstraction of the transformed language or from side effects. Abstraction means that there
ought to be a distinction between XML as output and transformation descriptions that operate on it. The ideal would be to obtain transform rules ``when \textit{Pattern} then \textit{Substitute}``.
Those disadvanteges hinder the validation of composed transformation rules because the rules become unreadable and complex. Side effects could cause assertion violations and even so the orthogonality of the transformation can not be guaranteed and finally the target document becomes falsely generated.\\

Nevertheless a logical approach must injure referential transparency due unifying terms. Side effects and predicates constrain semantics to be denoted not in a functional but in a predicative style. A logical
approach will also be of interest because no proper transformation language has been proposed yet. The non linear character predicates can be evaluated seems to be promising in reforming more compact transformations. Therefore it can be assumed a shorter and more abstract transformation is obtainable. The more abstract a transformation becomes the easier it gets to specify it. It can also be stated that using
a logical transformation approach simplifies the acquisition because queries to XML documents already are 
transformations. This way it is possible to avoid semantical anomalies such as in XSL-T that depends on
XPath as its query language for external XML sources but itself is not describable in XML. The result of
previously determined results is implicitly propagated to its descending child nodes. In conclusion the following threads are going to be researched:

\begin{itemize}
	\item Find a base of transform and query operators.
	\item Simplify transformations that need many queries to build similiar blocks into the target document.
	\item Investigate cases in which inverse predicates serve as validator.
	
	\item Push up limits from recent XPath and XSL-T reference implementations concerning syntactical constraints, e.g.	user defined functions, closed world and multi indices \cite{XPath,XSLT}.
\end{itemize}

The aim of this article is to define a logical language for XML processing based on Prolog and
to evaluate this logical transformation language (LTL) in comparison to the clear XML transformation language XSL-T. Primarily the comparison is oriented to user requirements formulated in today's XSL-T
forums. It is intended LTL is extensible as much as possible to break through the limited syntax of XSL-T.
The task restricts to XML documents, but essentially all hierarchical markup documents are affected.
Domains of a LTL are complex schema-transformations, text retrieval and the analysis of formal languages.

First work on logical transformations on XML documents has partially been done in \cite{Seipel}. Unfortunately
the described mismatch between query and target languages proceeds and most of defined predicates just reimplement a
functional framework attempt. Consequently predicates became tough combinators and the evaluation order got inflexible because profound language features of Prolog missed. Therefore combinators became complex, numerous, redundant and finally error-prone.

\section{Importance}

The way knowledge is represented and infered logical is one common approach. Considering Prolog as 
logical programming language transformation rules can directly be represented as Horn clauses.
  In distinction to functional programming languages \cite{Wadler,Novatchev,XQuery,Kiselyov} a premise might be partially undetermined. In fact a premise
predicate might contain symbols unified after depending conjunctions have been evaluated or even never.
Prolog makes use of side effects to unify terms. Although these anomalies are locally to each Horn clause
which makes Prolog attractive for componing. Summarizing Prolog knowledge can be represented by facts and Horn clauses.
This is essential comparing Prolog with XQuery which allows exclusively rigid FLOWR expressions. Shorter code affects
a better formalisation of the transforming semantics and allows simpler checkings against given specifications.

It is notable that APIs have been developed \cite{TuProlog} that enable bindings to but also from Java. Therefore numerous feasibilities result an integration of previous XML processors. The opposite direction permits sophisticated
algorithms in Java been rewritten in Prolog. That might be capable for descriptive transformation languages, business rules policies, multi channel publishing etc.

Starting with an appropriate data model for LTL, a rather simple derivation rules will be defined later on.

\section{Data Model}
Well-formed XML documents at least consist of element, text and attribute nodes (figure ~ \ref{fig:BNFDataModel}).
  It can be assumed that parsed and serialized documents are well-formed and canonized hence valid normalforms simplify
analysis of duplicates. Canonized documents describe documents which contain attributes that have been sorted in lexicographic ascending order by attribute name. Standard Prolog has no static type checking, so syntactical invalid transformations unfortunately can not be detected before serializing the target document. 
 Since Prolog does not offer pointer references it gets necessary to specify nodes completely as short as possible
in its full specification. Lists seem fairly to represent attribute and children sets. Only hedges or children sets
must satisfy a strict order. Analogous defined nodes could be rewritten in lists as it was done by \cite{Seipel}. But
this attempt lack on proving node's type and consequently does not bring any advantage. Despite an uniform syntax affected predicates would become even more complex because precise distinction between list members would be necessary.
 
\begin{figure}[ht]
\begin{center}
\begin{grammar}
<hedge> ::= $\varepsilon$ | <node> | <node> ',' <hedge>\\
<node> ::= <element> | <textnode> | <pi> | <comment>\\
<element> ::= 'element' '(' <name> ',' '[' <attributes> ']' ',' '[' <hedge> ']' ')'\\
<textnode> ::= 'text' '(' <text> ')'\\
<pi> ::= 'pi' '(' <text> ')'\\
<comment> ::= 'comment' '(' <text> ')'\\
<name> ::= <alpha> <name2>\\
<text> ::= <name2>\\
<name2> ::= $\varepsilon$ | <alphanum> <name2>\\
<alphanum> ::= <alpha> | '0' | .. | '9'\\
<alpha> ::= 'a' | .. | 'z'\\
<attributes> ::= $\varepsilon$ | <attribute> | <attribute> ',' <attributes>\\
<attribute> ::= '"'"' <name> '=' $''$ <text> $''$ '"'"'
\end{grammar}
\end{center}
\vspace{-0.5cm}
	\caption{Backus-Naur form of Prolog data model representing XML documents. The superior node in a valid XML document has to be an element node.}
	\label{fig:BNFDataModel}
	
\end{figure}

 In conclusion we get a more intuitive and compact representation that avoids mandatory writing of new combinators even for simple transformations. Furthermore there is no dependence on certain transformation orders due control can be influenced by individual predicates.\\

To describe any XML document it is enough to use element and text nodes.\\

This statement first induces that all other nodes - processing instruction and comment - can be expressed just by unambiguous element and text nodes (cmp. fig. \ref{fig:BNFDataModel}).

To show this it is sufficient to find some unambiguos encodings $\sigma,\tau$ and to expand the alphabet by some special symbols $\{\pi, \kappa\}$ with $\sigma: pi (\underline{text}) \mapsto \pi \cdot \underline{text}$ and $\tau: comment ( \underline{text} ) \mapsto \kappa \cdot \underline{text}$. The binary operation $\cdot$ concates two strings.

Further it states that attributes in element nodes can also be rewritten. Hence attribute nodes contain two
arguments it is obvious to use a tagged encoding such as element nodes. So an appropriate encoding would be:
$\alpha: \ '\underline{name} = ''\underline{text} '' \ ' \mapsto element(\underline{name},[],[text(\omega \cdot \underline{text})])$.
The element node of the right side specifies the upper element node, so it is necessary to agree upon inserting first
all attributes.

Such applying $\sigma,\pi,\alpha$ to a given XML document a XML document results that consists only of element and 
text nodes. This function obviously is injective. New node types are encoded in analogy. Nodes from different name spaces are handled as usual element nodes with a prefixed name $\square$.\\

Given a XML document $X$, $\forall u,v \in X: reachable(u,v)$ is satisfied.\\

As it was shown earlier it is sufficient to consider the simplified case. Further such
a top level element by induction represents a n-way tree. From the tree character of $X^{'}$ the lenght of any path in $X^{'}$ is bounded by the number of nodes decreased by one $\square$.\\

Combinators defined on XML documents can be eliminated in Prolog.\\

As seen in the previous statement the path can be build just by ascending and descending
relationships considering a bidirectional tree. The direction depends on the unification of unevaluated node constructors in Prolog. So, combinators are sufficient but are not necessary $\square$.\\
 
Prolog gives universal operators for transformations for free. Additional operators are not necessary from the computational view.\\

When Prolog's rules are interpreted as tables a relational algebra equivalent to Codd's algebra can be defined
as following. $R,S$ and $T$ describe any relations. 

\begin{tabular}{rl}

\textbf{Union} $T=R\cup S$: & \parbox[t]{10cm}{$t(x_{1},...,x_{m}):-r(x_{1},...,x_{m}).$\\
$t(y_{1},...,y_{n}):-s(y_{1},...,y_{n}).$}\\

\textbf{Difference} $T=R/S$: & \parbox[t]{5cm}{$t(x_{1},...,x_{n}):-r(x_{1},...,x_{n}),$}\\

                             & \parbox[t]{10cm}{$not(s(x_{1},...,x_{n})).$}\\

\textbf{Carthesian} $T=R\times S$: & \parbox[t]{10cm}{$t(x_{1},...,x_{m},y_{1},...,y_{n}):-$}\\
				   & \parbox[t]{10cm}{$r(x_{1},...,x_{m}),\\ s(y_{1},...,y_{n}).$}\\

\textbf{Projection} $T=\Pi_{S}(R)$: & \parbox[t]{10cm}{$t(s_{1},...,s_{n}):-r(x_{1},...,x_{m}).\\ \forall s_{i} \in \{x_{1},...,x_{m}\}$}\\

\textbf{Selection} $T=\sigma_{S}(R)$: & \parbox[t]{10cm}{$t(x_{1},...,x_{n}):-$}\\
				      & \parbox[t]{10cm}{$r(x_{1},...,x_{n}),s(x_{1},...,x_{n}).$}\\

\textbf{Renaming} $T=\rho_{S}(R)$: & \parbox[t]{10cm}{$t(x_{1},...,x_{n}):-r(x_{1},...,x_{n}).\square$}
\end{tabular}\\

Deletion, Insertion and Exchange of elements in a XML documents requires one step in Prolog.\\

These three operations correlate with a similar resulting document within a binary relation $\Psi$. Among
the arguments in the document unification in $\Psi$ matches exactly except the changed element, so the common unificator contains exactly one substitution hence hedge manipulation can be fulfilled in one step in terms in Prolog $\square$.\\

Since it is not necessary to fully specify the part that has not been changed in Prolog, the number of manipulations is bound by a constant. There is no need to reconstruct invariant parts like in XSL-T. Such manipulating operations significantly improve the performance of the transformation and are called \textit{non-monotonic} due they avoid permanent input and output transfers.

\section{Transformation Rules}
\label{SectTransformationRules}
The specification of the most important transform operators can be found in the appendix. Depending
on their arity the operators can be chained to a path expression analogous to XPath. In distinction to
XSL-T the object and meta languages match with each other. Nodes are constructed and queried just in
Prolog.

 The search strategy of Prolog changes the instantiation process of the target document such that alternatives
will only be cut manually. Otherwise combinatorical parts will cause an exponential amount of valid target documents even if the user is only interested in the first element of each result set. For instance, it can be assumed that the LTL would be exactly implemented as defined by inference rules in the appendix. Hence there is no need to stop derivation after first matching of the third (//) rule, the search for alternatives could be continued with the fourth (//) rule. In such a case the derivation of the template

\begin{verbatim}
template(element(top,_,[A,A]),[text(T)]):-
   A=element(a,_,_),transform(A//p#1,T).
\end{verbatim}

which is semantically similar to the XSL-T stylesheet

\begin{verbatim}
<xsl:template match="top[count(child::*)
  =2] and a[1] and a[2]">
      <xsl:text>
     	   <xsl:value-of select="//a//p"/>
      </xsl:text>      
</xsl:template>
\end{verbatim}

would return a multiset of text nodes. In comparison, the given XSL-T example would stop in every case even if
\texttt{value-of} throws an exception. In fact it is not necessary within the implementation to consider
alternatives after one matching succeeds, so red cuts were set. Further as it can be seen in the previous Prolog template it would be rather simple to specify excactly the hedge \texttt{[A,A,B,A]} in \texttt{top} for any element node \texttt{B}.

Delete operators as specified in the appendix tend to shorten the filling of target documents due reducing
operations that are not necassary through copying unmodified fragments. Hence in worst case the whole document
becomes empty and due to the undecidability of the halt problem, a trade-off concerning the usage of non-monotonic
and average operators has to be done by the user. Even in the worst case the problem complexity is restricted
by the XSL-T boundary. It should also be taken into account that deleting any element node by a recursive path
expression, as it was not done here, affects an ambiguous validation since direct relationships get lost.

\section{Comparison}

\textbf{Expressiveness.}
To compare Prolog with XSL-T it is useful to introduce in XPath equivalent operators. For control structures no special
operators were needed because they fit in Prolog's declarative recursion schema.

 As announced earlier compact node specification, comfortable aggregations, and the usage of \textit{normal} and non-monotonic predicates enables shortened notations. All except ascending operators navigate within a given root document.

The built-in data structures of Prolog as atoms, tupels, lists and symbols widen XSL-T's default element hedge type. In fact in XSL-T node expressions often have to be converted to desired types. Most conversions have been covered by the path language XPath. Despite this standard Prolog unfortunately has no static type system which significantly diminishes practise acceptance to application programmers. Prolog avoids unnamed predicates and
requieres predicates to be called with a complete parameter list. Predicates that were called with less arguments as allowed in XSL-T were not evaluated or were evaluated with a body associated different to the intented body hence scoping of previous definitions is permitted. In general it is rather difficult to define functions by the user for reuse in XSL-T except namespaces and it is even more difficult to find an adequate representation that narrows its mathematical description. So it is quite impossible to formulate an adequate attribute sorting due the restrictiveness and invariance of built-in functions of XPath in XSL-T version 1.0.\\[-0.1cm]

\textbf{Usability.}
The Well-Formedness of documents is not guaranteed neither by XSL-T nor by Prolog. So XSL-T can generate hedges instead
of one top level element node and affect an unsuitable output by choosing an incorrect output encoding. Prolog can also
be effected by similar effects, but hedges on top niveau will overwrite previous alternatives successively.
To investigate the behavior of Prolog in comparison to XSL-T a measurement of common characteristics was done. Therefore
so called \textit{H\aa lstead metrics} were used \cite{Halstead}. Originally they are applied to procedural programming
languages. But LTL and XSL-T also have command tags which can be interpreted as operators $\eta_{1}$ and operands $\eta_{2}$ that are described by tag attributes. In more detail tag attributes could also be parsed but this should be avoided so far since results shall not be fixed to one special query language but rather to a transformation language. H\aa lstead metrics were primary not designed for markup processing, but hence markup is transformed on a higher level of abstraction in a command-like notation, it is worth to notice quantitative results. Used H\aa lstead metrics include measured program length $N$, theoretical program length $N_{T}$ and intellectual level $L$ and language abstraction niveau $\lambda$ that all base only on $\eta_{1}$, $\eta_{2}$ and an average estimation for failure made by an average-talent application programmer (see appendix). The experimental base counts about 70 different examples that mostly were taken from various XSL-T tutorials and completed by some individual test cases. Each example had two realizations in LTL, one that used exclusively predicates without templates and one that used templates as far as possible.

In common Prolog schema were about 50\% shorter than XSL-T variants according to Lines of Code, $\lambda$ and $\Delta_{N}=\left\|N_{T}-N \right\|$. This means Prolog programs are significantly more readable than XSL-T stylesheets.
Also $N_{T} : N$ and $\eta_{1} : \eta_{2}$ showed that Prolog is approximately 30\% more functional than XSL-T because their relations are more balanced. It can also be deduced that code becomes more difficult in Prolog because of the higher language abstraction niveau. Explicite traverses were almost always shorter than the variants using templates.\\[-0.1cm]

\textbf{Design.}
XSL-T lacks of closedness. There is no proper method in version 1.0 to define neither native XPath functions nor
adequate notation in XSL-T. Contrary, Prolog permits in-script definitions that are appropriate and modularized among libraries or scoped namespaces. Escpecially in layered architectures for Prolog-Java APIs Java functionality and universality can comfortable be plugged into existing higher level code.\\[-0.1cm]

\textbf{Implementation.}
An implementation has been realised using tuProlog \cite{TuProlog} and Java for String operations.
The transformation is a query to the existing knowledge base which consists at least of the basic 
operators defined in the appendix. For example the user is allowed to write some transformation rules
in Prolog as it was done in section \ref{SectTransformationRules}. Some template will be processed as soon as some node in the input XML file matches with the first parameter \texttt{element(top,\_,[A,A])} where \texttt{A} is the element node \texttt{a} specified by \texttt{A=element(a,\_,\_)}. As result the first occurence of a descended element node \texttt{p} is found and its associated text node is returned as secound argument in the template rule. Beneath nodes of top will not be traversed if it is not specified explicitly by new \texttt{template} calls in a template. The traversing continues with sibling nodes relatively to the matched node.

\section{Conclusion: Error Detection and Correction}

Error detection, diagnostic and correction are one of the most important tasks of formal language analysis. For performing Prolog-based transformations of XML document the user has to write appropriate programs in Prolog. It can cause errors in the target documents owing to lack of experience of the user. There are two approaches to diminish the probability of this nuisance. The first one is to design and implement a special interface that will be able to save the user from programming necessary transformations in Prolog. It is desirable that such interface would be easy for application and not to require knowledge of Prolog. Naturally some converter should be designed for transformation of texts formed in interface into appropriate Prolog programs. The second approach is to analyze the target document in order to detect errors. We suppose the best methods of detecting errors are global ones as they find minimum of errors and secure syntactic validity of the document after their correction. Now it is difficult to prefer one of these approaches. Both require further investigations.

\appendix

\subsection{Metrics}

\begin{tabular}{l}
$N_{T}=\eta_{1}ld(\eta_{1}) + \eta_{2}ld(\eta_{2})$\\
$\lambda=V*L$\\
$V=Nld(\eta)$
\end{tabular}

\subsection{Aggregation Operators}

\parbox[t]{5cm}{\inference[$last_{E}$]{E=element(\_,\_,[\cdots, C])}{C}}\\\\

\parbox[t]{5cm}{\inference[$count_{E}$]{E=element(\_,\_,[C_{1},...,C_{n}])}{n}}\\\\

\parbox[t]{5cm}{\inference[$lvl_{\vec{X}}$]{X_{0}=element(\_,\_,[\cdots,C_{0,i_{0}},\cdots]), C_{0,i_{0}}=X_{1}\\ X_{1}=element(\_,\_,[\cdots,C_{1,i_{1}},\cdots]), C_{1,i_{1}}=X_{2}\\ \vdots \\ X_{n-1}=element(\_,\_,[\cdots,C_{n-1,i_{n-1}},\cdots])\\ C_{n-1,i_{n-1}}=X_{n}=element(\_,\_,\_)}{[i_{0},i_{1},\cdots,i_{n-1}]}}\\\\

\subsection{Non monotonic Operators}

\parbox[t]{5cm}{\inference[$copy_{X}$]{X=element(\_,\_,\_)}{X}}\\\\
	
\parbox[t]{5cm}{\inference[$copy\_of_{X}$]{X=element(N,A,C)}{element(N,A,[])}}\\\\

\parbox[t]{5cm}{\inference[$remEl_{E,N}$]{E=element(Name,A,C) & \\ \texttt{append}(Pre,[element(N,\_,\_)|Post],C)\\\texttt{append}(Pre,Post,C2)}{element(Name,A,C2)}}\\\\
	
\parbox[t]{5cm}{\inference[$rem_{E,N}$]{E=element(Name,A,C)\\ \texttt{append}(Pre,[N|Post],C)\\ \texttt{append}(Pre,Post,C2)}{element(Name,A,C2)}}\\\\

\subsection{XPath equivalent Operators}

\parbox[t]{5cm}{\inference[$/_{E,N}$ ]{E=element(\_,\_,[\cdots,element(N,A,C),\cdots])}{element(N,A,C)}}\\\\

\parbox[t]{5cm}{\inference[$?_{E}$]{E=pi($''X''$)}{$'X'$}}\\\\

\parbox[t]{5cm}{\inference[$//_{E,N}$]{E=element(N,\_,\_)}{E}}\\\\

\parbox[t]{5cm}{\inference[$//_{E,N\neq X}$]{E=element(X,\_,[])}{$fail$}}\\\\

\parbox[t]{5cm}{\inference[$\#_{E}$]{E=text($''X''$)}{$'X'$}}\\\\
  
\parbox[t]{5cm}{\inference[$//_{E,N\neq X}$]{E=element(X,\_,[H|T])}{H$//$N}}\\\\

\parbox[t]{5cm}{\inference[$id_{E,Val}$]{E=element(\_,[],\_)}{$fail$}}\\\\

\parbox[t]{5cm}{\inference[$//_{E,N\neq X}$]{E=element(X,\_,[H|T])}{element(X,\_,T)$//$N}}\\\\

\parbox[t]{5cm}{\inference[$@_{E,Att}$]{E=element(\_,[],\_)}{$fail$}}\\\\

\parbox[t]{5cm}{\inference[$@_{E,Att}$]{E=element(\_,[\cdots,$'Att=''Val'' '$,\cdots],\_)}{$'Val'$}}\\\\
  
\parbox[t]{5cm}{\inference[$id_{E,Val}$]{E=element(\_,[\cdots,$'Att=''Val'' '$,\cdots],\_)}{$'Att'$}}\\\\

\parbox[t]{5cm}{\inference[$descendant_{E}$]{E=element(\_,\_,[])}{$fail$}}\\\\

\parbox[t]{5cm}{\inference[$descendant_{E}$]{E=element(\_,\_,\_)\\ E2=E/\_}{$E2// \_$}}\\\\
	
\parbox[t]{5cm}{\inference[$child_{E,C}$]{E=element(\_,\_,[\cdots, C, \cdots])}{C}}\\\\



\begin{thebibliography}{9}
  
  \bibitem{Seipel}
  \noindent Dietmar Seipel. Processing XML Documents in PROLOG. Proc. $17^{th}$ Workshop on Logic Programming. WLP2002.
  
  \bibitem{Janssen} Wim Janssen, Alexandr Korlyukov, Jan Van den Bussche.
				  On the tree-transformation power of XSLT.
				  arXiv.org 2006\\
				  \textit{http://arxiv.org/pdf/cs.PL/0603028}

  \bibitem{Novatchev} Dimitre Novatchev.
				 The Functional Programming Language XSLT - A proof through examples.
				 November 2001\\
				 \textit{http://www.topxml.com/xsl/articles/fp}
				 
	\bibitem{Kiselyov} Oleg Kiselyov, Shriram Krishnamurthi.
				 SXSLT: Manipulation Language for XML. Proc. Fifth Intl. Sym. Practical Aspects of Declarative Languages. PADL 2003\\
				 \textit{http://www.cs.brown.edu/\textasciitilde sk/Publications/Papers/Published/kk-sxslt}
				 
	\bibitem{Wadler} Phil Wadler.
				A formal semantics of patterns in XSLT	Markup Technologies.
				CiteSeer 1999
				\textit{http://citeseer.ist.psu.edu/204315.html}
					
	\bibitem{Xact} Xact XML-Transformation Framework for Java.\\
				 \textit{http://www.brics.dk/Xact}
				 
	\bibitem{Halstead} H\aa lstead, Maurice H. Elements of Software Science, Operating, and Programming Systems Series Volume 7. New York, NY: Elsevier, 1977.
				 
  \bibitem{TuProlog} Enrico Denti, Andrea Omicini, Alessandro Ricci.
tuProlog: A Light-weight Prolog for Internet Applications and Infrastructures - Practical Aspects of Declarative Languages. 3rd International Symposium (PADL'01). Springer-Verlag 2001
				 
  \bibitem{XQuery} XQuery 1.0 Specifikation, 2006.\\
				 \textit{http://www.w3c.org/TR/xquery}
           
  \bibitem{XSLT} XSLT 1.0 Specification, 1999.\\
				 \textit{http://www.w3.org/TR/xslt}
         
	\bibitem{XPath} XPath 1.0 Specification, 1999.\\
				 \textit{http://www.w3c.org/TR/xpath}  
				 
\end{thebibliography}
\end{document}